From rvl1@cornell.edu  Wed Jun 25 16:29:29 1997
Received: from postoffice.mail.cornell.edu (POSTOFFICE.MAIL.CORNELL.EDU [132.236.56.7]) by spacenet (8.6.12/8.6.12) with ESMTP id QAA00298 for <scheel@spacenet.tn.cornell.edu>; Wed, 25 Jun 1997 16:29:29 -0400
\documentstyle[aas2pp4,tighten]{article}
\begin{document}


\title{
Influence of Ohmic Heating on Advection-Dominated
Accretion Flows }

\author{G.S. Bisnovatyi-Kogan}
\affil{Space Research Institute,
Russian Acadamy of Sciences,
Moscow, Russia; gkogan@mx.iki.rssi.ru}

\author{R.V.E. Lovelace}
\affil{Department of Astronomy,
Cornell University, Ithaca, NY 14853-6801;
rvl1@cornell.edu}

\begin{abstract}

Advection-dominated, high-temperature,
quasi-spherical accretion
flow onto a compact object,
recently considered by a number of authors,
assume that the dissipation of turbulent energy
of the flow heats the ions and
that
the dissipated energy is advected inward.
 It is suggested that the
efficiency of conversion
of accretion energy to radiation
can be very much smaller than unity.
 However, it is
likely that the flows have an equipartition magnetic
field with the result that dissipation of magnetic
energy at a rate comparable to that for the
turbulence must occur by Ohmic heating.
 We argue that this heating occurs as a
result of plasma instabilities and that the
relevant instabilities are current driven
in response to the strong electric fields
parallel to the magnetic field.
 We argue further that
these instabilities heat
predominantly the electrons.
 We conclude that the efficiency
of conversion of accretion energy to
radiation can be much smaller than
unity only for the unlikely condition
that the Ohmic heating of the electrons
is negligible.
\end{abstract}

\keywords{accretion,
accretion disks---galaxies:
active---plasmas---magnetic
fields---stars: magnetic
fields---X-rays: stars}

\section{Introduction}

	Advection-dominated accretion flows
have been intensely studied during the past
several years (for example, Narayan and Yi 1994;
Narayan and Yi 1995;
Abramowicz
et al. 1995; Nakamura et al. 1996;
Chakrabarti 1996).
 The basic dynamical
equations for accretion disks including the
advection of entropy were first
discussed by Paczy{\'n}ski
and Bisnovatyi-Kogan (1981)
and Muchotrzeb
and Paczy{\'n}ski (1982).
 In contrast with
the widely applied theory of thin
accretion disks of Shakura (1973)
and Shakura and
Sunyaev (1973)
where the disk material
cools efficiently by local
radiation of viscously generated
energy, the advection-dominated
accretion flows of Narayan and Yi {\it
assume} that
the viscous dissipation heats the ions
and that a constant
fraction $f$ of this dissipated
energy is advected inward and that
the fraction $1-f$ is locally radiated.
 The further assumption that
the energy exchange between
ions and electrons is by Coulomb scattering
leads to conditions with the
ion temperature $T_i$ much
larger than the electron
temperature $T_e$ so that the
cooling is inefficient.
 [A recent paper by Esin et al. (1996)
treats advection dominated
accretion flows assuming
$T_i=T_e$.]
 The {\it radiative efficiency},
the power
output in radiation divided by $\dot M c^2$
(with $\dot M$ is the mass accretion rate),
is found to be very small compared
with unity.
 The
advection-dominated
accretion flows tend to
be quasi-spherical and
optically thin (except for cyclotron
radiation as discussed below) with
radial inflow speed
$v_r \approx - \alpha v_{K}$,
azimuthal speed $v_\phi\approx
{\rm const.}v_{K} \ll v_{K}$,
ion thermal speed $c_{si} \approx {\rm const.}
v_{K} \sim v_{K}$ (Narayan and Yi 1995),
where $v_{K}\equiv (GM/r)^{1\over2}$ is
the Kepler speed and $\alpha$
is the dimensionless
viscosity parameter of
Shakura (1973)
usually
assumed to be in the range $10^{-3} -
1$.

In \S 2 we discuss magnetized accretion flows
and the importance of Ohmic dissipation in
addition to the earlier considered viscous
dissipation.  We argue that the Ohmic heating
is due to plasma instabilities which heat
the electrons.  In \S 3 we treat a
model for the radial variation of electron and
ion temperatures
assuming that a fraction $g$ of the
dissipated energy goes into heating
the electrons and a fraction $(1-g)$ goes
into heating the ions.
The electrons cool by bremsstrahlung
and cyclotron radiation
and exchange energy with ions by Coulomb
collisions.
In \S 4 we discuss conclusions of
this work

\section{Accretion Flows with $\bf B$ Field}

	In quasi-spherical accretion
onto a compact object of mass $M$
of Schwarzschild radius
$r_S \equiv 2GM/c^2$  (for a black hole)
the accreting matter is likely to be
permeated by a magnetic field ${\bf B} ({\bf r}, t)$.
 Typically the accreting
matter is ionized and
consequently highly conducting
with the result that
the magnetic field is frozen into the
flow.
 One result of this is that $|B_r| \propto r^{-2}$ .
Thus the magnetic
energy-density varies as
${\cal E}_{mag} = {\bf B}^2/8\pi \propto r^{-4}$.
On the other hand the kinetic energy-density
varies as ${\cal E}_{kin} =
\rho{\bf v}^2/2  \propto r^{-{5\over 2}}$.
Thus one can expect that
equipartition between magnetic and
kinetic energy-densities occurs in the flow at a
large distance $r=r_{equi} \gg r_S$ (Shvartsman 1971) and
that it is maintained for smaller $r$.
Further accretion for $r<r_{equi}$ is
possible only if magnetic flux is destroyed
by reconnection and the magnetic
energy ${\cal E}_{mag}$ is dissipated.
The dissipation of magnetic energy was first
taken into account by Bisnovatyi-Kogan
and Ruzmaikin (1974) who showed
that accretion for conditions
of equipartition
(${\cal E}_{mag}\sim {\cal E}_{kin}$)
is accompanied by the dissipation of
magnetic energy into heat with
entropy $s$ (per unit mass)
production rate $\rho T(ds/dr) =
-3{\bf B}^2/(16\pi r)$.
We point out that the Ohmic
dissipation of the magnetic energy
is an important, possibly
dominant heating process in
advection-dominated accretion flows with
${\cal E}_{mag} \sim {\cal E}_{kin}$.
In this regard note that although Narayan and Yi (1995)
assume an equipartition magnetic field, they do
not consider the Ohmic heating.

	The basic equations for
accretion flows with ${\cal E}_{mag} \sim {\cal E}_{kin}$  are
$$\left({\partial \over \partial t}+{\bf v \cdot \nabla}\right) {\bf v}=
-{1\over\rho}{\bf \nabla}p +{\bf g} +{1\over \rho c}{\bf J \times B}
+ \nu_m {\nabla^2}{\bf v}~
\eqno(1a)
$$
$${\partial {\bf B}\over \partial t} =
{\bf \nabla \times (v\times B)} + \eta_m \nabla^2 {\bf B}~,
\eqno(1b)$$
where ${\bf v} ({\bf r}, t)$ is the
flow velocity, $p({\bf r},t)$ the pressure,
${\bf g} = -{\bf \nabla}(G M/|{\bf r}|)$
the gravitational acceleration, $\nu_m$ the {\it
microscopic} kinematic viscosity coefficient,
and $\eta_m$ the {\it microscopic} magnetic diffusivity.

	It is well known that the microscopic classical
transport coefficients $\nu_m$ and $\eta_m$
are much too small to directly
influence on the macroscopic
flow $\bf v$ and magnetic field $\bf B$
evolution.  For example, for conditions
pertinent to a flow onto a massive black hole,
$n \sim 10^{12}{\rm cm}^{-3}$, $T_i\sim 10^{12}{\rm K}$,
and $B\sim 10^4{\rm G}$ for $r\sim r_S$,
the Reynolds number for the
flow $Re_v = r|{\bf v}|/\nu_m \sim 10^{24}$, where
$\nu_m \sim r_{gi}^2/ \tau_{ii}$
is the viscosity appropriate
for a tangled magnetic field (Braginskii 1965;
Paczy{\'n}ski 1978), and
where
$r_{gi}\sim 10^2{\rm cm}$ is the
ion gyro-radius and $\tau_{ii}\sim 10^6{\rm s} $ is the ion-ion
Coulomb scattering time, and $\omega_{ci}\tau_{ii} \gg 1$ with
$\omega_{ci}\sim 10^8/s$ the ion cyclotron frequency.
[Under some conditions it
is possible that $\nu_m$
is larger than
$r_{gi}^2/ \tau_{ii}$ as
discussed by Subramanian, Becker,
and Kafatos (1996)].
The magnetic Reynolds number
$Re_B = r|{\bf v}|/\eta_m \sim 10^{27}$, where
$\eta_m = c^2/(4\pi \sigma_S)$ with $\sigma_S$ the
Spitzer conductivity.

It was proposed by Shakura
(1973) that accretion flows
are in general turbulent
and that roughly equations (1)
should be taken with turbulent
transport coefficients $\nu_t$
and $\eta_{t}$  replacing the
microscopic coefficients,
and with ${\bf v} \rightarrow \bar{\bf v}$
and ${\bf B} \rightarrow \bar {\bf B}$
interpreted as {\it
mean fields}.  The turbulent viscosity
has a crucial role in thin
Keplerian disks where it provides a
mechanism for the outward transport of
angular momentum.  According to
Shakura (1973), $\nu_t = \alpha c_{si} H$,
where $\alpha = {\rm const.}$ is the above-mentioned
dimensionless viscosity parameter, $c_{si}$ is
the ion sound speed, and $H$ is the
half-thickness of the disk which is
the {\it outer-scale} of the turbulence.
Note that for an advection-dominated
accretion flow, $H\sim r$.
 The shear stress
in a magnetized accretion flow, which causes
outflow of the angular momentum, appears in large
part be due to magnetic stress (Eardley and Lightman 1975;
Brandenberg et al. 1995; Hawley, Gammie, and Balbus 1995).
Bisnovatyi-Kogan and Ruzmaikin (1976)
argued that $\eta_t \sim \nu_t$.  The turbulent
diffusivity
will have a crucial role in dissipating
the magnetic energy in advection-dominated flows.
In addition to $\nu_t$ and $\eta_t$, there will
be a turbulent transport coefficient $\alpha_h$
(with units of $cm/s$) associated with the
helicity of the turbulence in a rotating
accretion flow (see, for example, Ruzmaikin, Shukurov,
and Sokoloff 1988).

	Neglecting for the moment the possible
difference between $T_e$ and $T_i$ and the
radiative energy losses, energy conservation for the
accretion flow can be
expressed in terms of the mean
fields as
$$ \rho T {ds \over dt} = {1\over 2} \rho \nu_t
\left(\bar v_{i,j} +\bar v_{j,i}-
{2\over3}\delta_{ij}\bar v_{k,k}\right)^2
$$
$$+{1\over 4\pi}\eta_t\left({\bf \nabla \times \bar B}\right)^2~,
\eqno(2)$$
where $s$ is the entropy per unit mass.  The first term
on the right hand side of (2) represents the viscous
dissipation or heating of the plasma, and the
second term the Ohmic dissipation.  The two
terms are of comparable magnitude for an
accretion flow with ${\cal E}_{mag} \sim {\cal E}_{kin}$
and $\nu_t \sim \eta_t$.

However, equation (2)
says nothing about the {\it actual}
microscopic dissipation of energy in the
plasma.  Rather, it expresses the loss of energy
from the outer-scale ($\sim r$ or $H$ if $H \ll r$) of the
flow $\bf v$ and from the
$\bf B$ field by the nonlinear processes
implicit in equations (1) and the presumed
Kolmogorov
cascade of this energy to smaller scale
eddies and field structures of the flow.
The turbulence may be characterized by
wavenumber-frequency ensemble averaged
spectra $\langle{\bf v}^2_{{\bf k}\omega}\rangle$ and
$\langle{\bf B}^2_{{\bf k}\omega}\rangle$, where the
wavenumber ranges from the small value
corresponding to the mentioned outer
scale $k_{min}\sim r^{-1}$ to some much
larger value $k_{max} \gg k_{min}$.
 The
conventional Kolmogorov description has
a dissipation scale corresponding to
$k_{max} \sim (Re)^{3\over 4} k_{min}$
which corresponds to an unphysically
small length scale using either $Re_v$
or $Re_B$.
 Thus, the actual dissipation must be due plasma
instabilities.

 The relevant plasma instabilities
are probably current driven in response to the
large mean electric field, $\bar{\bf E}=
-\bar{\bf v} \times \bar{\bf B}/c -\alpha_h \bar {\bf B}/c
+\eta_t {\bf \nabla}\times \bar {\bf B}/c$, which in
general has a significant component parallel to ${\bf B}$.
 It is unclear to us why current driven instabilities
resulting from $E_\parallel$ were not considered by
Begelman and Chiueh (1988).
 The typical electric field $|{\bf E}| \sim 10^6 ~V/cm$
(for $r\sim r_S$) is much larger than the Dreicer electric
field for electron runaway (Parail and Pogutse 1965),
$E_D = 4\pi e^3(n_e/kT_e)\ell n
\Lambda \sim 10^{-4}~ V/cm$ for $T_e\sim 10^9K$,
where $n_e$ is the electron density.
Thus the electrons will runaway.
An electron becomes relativistic in a distance of
travel of $\sim 1~cm$ which is comparable to the
electron gyro radius.  The drift speed of the
electrons parallel to ${\bf B}$ will be sufficient
to give rise to streaming instability (Parail and
Pogutse 1965).
 Streaming instability will occur if the
electron drift velocity is larger than
the ion thermal speed.
 In contrast with the ions, the travel distance
for a proton to become relativistic is $\sim 10^3 ~cm$.
 However, acceleration of protons
parallel to the magnetic
field is strongly
suppressed by scattering by
magnetic fluctuations (Alfv\'en waves)
with wavelengths of the
order of the proton gyro radius which are generated
by the proton streaming (Kulsrud and Pearce 1969).
For these reasons we believe that most of
the free energy driving the instability goes
into heating the electrons.
However, we also consider the case where
a fraction $g$ of the dissipated
energy goes into heating
the electrons and $(1-g)$ goes into heating
the ions.
We illustrate the behavior in this case with
the following simple model.

\section{Model}
We generalize equation (2) by taking into
account (a) that $T_i$ and $T_e$ may differ
with energy exchange between ions and electrons
by Coulomb collisions,
(b) that the Ohmic plus viscous
dissipation heats
electrons and ions as discussed below,
and (c) that the main energy
loss is from optically thin bremsstrahlung and
optically thick cyclotron emission.
 Note that the thickness of the flow $H/r$ is not restricted.
Note also that in contrast with Narayan and Yi (1995), no
assumption is made that a constant fraction $f$ of the dissipated
energy is advected inward.
Hence
$${3\over2}{dT_i\over dt}-{T_i\over\rho}{d\rho\over dt}=
(1-g){\cal H}-\nu_{ie}(T_i-T_e)~,
\eqno(3a)$$
$${3\over2}{dT_e\over dt}-{T_e\over\rho}{d\rho\over dt}=
g{\cal H}-{\cal C}_{brem}-{\cal C}_{cyc}+\nu_{ie}(T_i-T_e)~,
\eqno(3b)$$
where $g\le 1$ is the fraction
of the Ohmic plus viscous dissipation which
goes into heating the electrons.
 We assume
$g={\rm const.}$ which we view as more
physically plausible than the assumpition
that $f={\rm const.}$ of Narayan and Yi.
 For simplicity of the formulae we assume
$T_i<m_ic^2$ and $T_e<m_ec^2$,
where $T_i$ and $T_e$ are measured in ergs.
Here,
$$\nu_{ie} \approx
{4(2\pi)^{1\over2}n e^4 \over m_im_e}
\left({T_e\over m_e}+{T_i\over m_i}\right)^{-{3\over2}}
{\ell}n\Lambda
$$
is the ion-electron energy exchange rate
with $\ell n \Lambda={\cal O}(20)$ the
Coulomb logarithm (Spitzer 1940);
${\cal H} \approx (9/4) m_i \alpha
(c_{si}/v_K)^2 $ $ v_K^3{\cal J}/r$
is the heating rate
per ion with ${\cal J}=1-(r_S/r)^{1\over 2}$;
${\cal C}_{brem} \approx n\sigma_T \alpha_f$ $
m_ec^3(T_e/m_ec^2)^{1\over2}$ is
the bremsstrahlung cooling rate per
electron with
$n$ the electron or ion density,
$\sigma_T$ the Thomson cross section, and
$\alpha_f$ the fine structure constant;
and ${\cal C}_{cyc} \approx T_e \omega_{ce}^3
{\cal M}_c^3/(8\pi^3 n c^2 r)$ is
the self-absorbed cyclotron
radiation cooling rate per electron,
with ${\cal M}_c \gg 1$ the cut-off
harmonic number of the cyclotron radiation below which
the radiation is self-absorbed (Trubnikov 1958).
  For ${\cal M}_c \gg(2/9)\mu \gg 1$,
with $\mu \equiv m_e c^2/T_e$,
Trubnikov's analysis gives
${\cal M}_c \approx (2\mu/9)(1+\ell n({\cal D})/\mu)^3$, where
${\cal D}\approx \omega_{pe}^2 r/
(c\omega_{ce} {\cal M}_c)$, with
$\omega_{pe}$ and $\omega_{ce}$
the electron plasma and
cyclotron frequencies respectively.
Trubnikov's expression for ${\cal C}_{cyc}$
is similar to that of
Narayan and Yi (1995).

 It is useful to rewrite equations (3) in dimensionless form.
Note that $d/dt= v_r(d/dr)$ with $v_r=-(3/2)\alpha \hat T_i v_K$,
and that $H/r = \hat T_i^{1\over2}$,
number density of electrons or ions $n=\dot M/(6\pi\alpha
m_ir^2 \hat T_i^{3\over2} v_K)$, mass density
$\rho = nm_i$, magnetic field $B=[2\dot M v_K/
(3\alpha r^2 \hat T_i^{3\over2})]^{1\over2}$,
where $\hat T_i \equiv T_i/T_v$
with $T_v\equiv GMm_i/r$ the
virial temperature.
 We also normalize the electron temperature with
the same $T_v$, $\hat T_e \equiv T_e/T_v$.
 Equations (3) become
$${d \hat T_i \over d \hat r} =
-(1-g)\hat{\cal H}+\hat A (\hat T_i -\hat T_e)~,
\eqno(4a)$$
$${d \hat T_e \over d\hat r}=
-[(2+\zeta)g-\zeta]\hat{\cal H} +\hat{\cal C}_{brem}+\hat{\cal C}_{cyc}
$$
$$-(2+\zeta) \hat A(\hat T_i-\hat T_e)~,
\eqno(4b)$$
where $\hat r \equiv r/r_S$, with $r_S$ the Schwarzschild radius,
$\zeta \equiv \hat T_e/\hat T_i$, and
where
$$\hat A \approx {8 \over 9 \pi^{1\over 2}\alpha^2}\left({m_e\over m_i}\right)
\left({\dot M c^2 \over L_E}\right)
\left(\hat T_i+ {m_i\over m_e}\hat T_e\right)^{-{3\over 2}}\times
$$
$$
{\hat r^{1\over 2}\over \hat T_i^{5\over 2} }~\ell n\Lambda~,
\eqno(4c)$$
$$ \hat {\cal H} \approx {{\cal J}\over 2\hat
r}~,~~~~~~~~~~~~~~~~~~~~~~~~~~~~~~~~~~~~~~~~~
\eqno(4d)$$
$$\hat{\cal C}_{brem}\approx{2^{1\over 2}8\alpha_f \over 27 \alpha^2 }
\left({m_e\over m_i}\right)^{1\over 2}\left({\dot M c^2\over L_E}\right)
{\hat T_e^{1\over 2} \over \hat r^{1\over 2} \hat T_i^{5\over 2}}~,~~~~~
\eqno(4e)$$
$$\hat {\cal C}_{cyc}\approx{ 1\over 9\pi^2 2^{1\over 4}\alpha^{3\over2}}
\left[\left({m_i \over m_e}\right)^3\left({r_e\over r_S}\right)
\left({\dot M c^2\over  L_E}\right)\right]^{1\over2} \times$$
$$\left({\hat T_e \over \hat T_i^{7\over 4}}\right)
{{\cal M}_c^3 \over \hat r^{11\over 4}}~,
\eqno(4f)$$
where $L_E \equiv 4\pi GM m_ic/\sigma_T$
is the Eddington luminosity,
and $r_e \equiv e^2/(m_ec^2)$ is the
classical radius of the electron.
 The terms $d\hat T_i/d\hat r$ and $d\hat T_e/d\hat r$ in
equations (4) describe the advection of energy by the flow.
 Apart from the
cyclotron cooling the different terms depend only
on $\alpha$ and $\dot M c^2/L_E$.
 The cyclotron
cooling is relatively more important for accretion
onto a stellar mass object than for accretion onto
a massive black hole.
 The assumed condition for optically thin
bremsstrahlung radiation requires
$(\dot M c^2/\alpha L_E)\hat r ^{-{1\over 2}} <1$
for $\hat T_i = {\cal O}(1)$.

 We have solved equations (4)
starting from different given `initial'
values of $\hat T_i$ and $\hat T_e $ at
large $\hat r = 10^3$, different accretion
rates $\dot M c^2 = (0.01-1)L_E$,
values of $\alpha$, and values of $g=0-1$,
and integrating inward.
 For the accretion rates where
advection-dominated flows are suggested to
occur (Narayan and Yi 1995),
$\dot M c^2  \leq 0.1 L_E$ for $\alpha=0.1$,
we find that the
the scaled ion temperature
$\hat T_i$ remains almost
constant, whereas the scaled
electron temperature $\hat T_e$
decreases rapidly as $\hat r$ decreases
from $10^3$.
 In this limit, the Coulomb
energy exchange between
ions and electrons is negligible.
 The advection terms
on the left-hand-side of equation (3b) are
also negligible.
Consequently, the Ohmic heating of the electrons
$g {\cal H}$ goes into radiation, mainly cyclotron
radiation;  that is,
$g{\cal H} \approx {\cal C}_{cyc}$.   The
total radiation is the volume integral of
$g{\cal H} n$ which gives $gGM\dot M/(2 r_i)$,
where $r_i$ is the inner radius of the flow.
Thus, the radiative efficiency is
reduced by a factor of $g$
from that of a thin disk with
$\hat T_i = \hat T_e \ll 1$ which is the
volume integral of ${\cal H}n$.
 This efficiency can
be very small compared with unity only
if $g$ is very small compared with unity.

\section{Conclusions}

This work considers magnetized
advection-domin-ated
accretion flows where the magnetic field is
in equipartition with the turbulent
motions of the flow (Shvartsman 1971).
 The magnetic energy density of the flow
must be dissipated by Ohmic heating with
a rate comparable to that of the viscous
dissipation (Bisnovatyi-Kogan and Ruzmaikin
1974).
 We argue that the Ohmic and viscous
dissipation must occur as a result of
plasma instabilities.
 Further, we argue
that the instabilities are likely to
be current driven in response to the
electric field (associated with the
turbulent motion) which has a significant
component parallel to the magnetic field.
 These instabilities are likely to heat
mainly the electrons.
 We have
analysed a model for the
radial variation of the electron and
ion temperatures assuming
that a constant fraction $g$
of the viscous plus Ohmic heating
goes into heating the electrons and
a fraction $(1-g)$ goes into heating
the ions.
 In contrast with Narayan and
Yi (1995), we do {\ it not}
assume that a constant
fraction $f$ of the dissipated energy is
advected inward by the flow.
 The electrons cool by
bremsstrahlung
and cyclotron radiation
and exchange energy with the ions by
Coulomb collisions.
 At large accretion rates $\dot M$,
Coulomb collisions act to give
$T_i \approx T_e$, high radiative
efficiency, and geometrically-thin,
optically-thick
disk accretion.
 For small accretion rates,
where advection-dominated accretion
flows are suggeested to occur,
and only Coulomb
energy exchange between ions and
electrons, a regime of optically thin
accretion flows
with a large difference between ion and
electron temperatures ($T_e \ll T_i$) exists
(Shapiro, Lightman, \& Eardley 1976).
 Here, we emphasize that
the accretion flow properties
depend critically on the Ohmic
heating of the electrons.
 For small accretion rates where
the electron temperature is much
less than the ion temperature, we
show that the Ohmic heating of the
electrons gives a radiative efficiency
which is reduced by a factor of $g$
from that for a thin disk.
 Thus, the tiny radiative
efficiencies ($<10^{-3}$) found by
Narayan and Yi (1995) correspond to
tiny values of $g$ which are unlikely
for the reasons discussed in \S2.

 Plasma instabilities due to electron-ion
streaming (for electron drift
velocity larger than the ion thermal
speed) may greatly enhance the
energy exchange between ions and
electrons.
In this case the
two-temperature regime disappears,
the ion and electron temperatures collapse
to small values,
$\hat T_{i,e} \ll 1$,
and the disk is geometrically thin.
That is,
advection-dominated accretion flows
do not occur (Fabian \& Rees 1995).

\acknowledgements{
We thank Drs. M.M. Romanova and
H.H. Fleischmann for
valuable discussions.
This work was supported by
NSF grant AST-9320068 and a grant
from the CRDF Foundation.  The work of
GBK was also supported by Russian
Fundametal Research Foundation
grant No. 96-02-16553.  The work of
RVEL was also supported by NASA
grant NAGW 2293.}

\end{document}